\documentclass[twocolumn,superscriptaddress,showpacs,amsmath,amssymb,pra,longbibliography]{revtex4-1}

\usepackage[pdftex]{graphicx} 
\usepackage{dcolumn}  
\usepackage{bm}       
\usepackage[usenames,dvipsnames]{color}
\definecolor{URLCOL}{rgb}{0,0.52,0.83} 
\definecolor{LINKCOL}{rgb}{0.05,0.5,0} 
\definecolor{orange}{rgb}{0.6,0.3,0} 
\definecolor{CITECOL}{rgb}{0.25,0,0.48} 
\usepackage{epstopdf}

\usepackage[pdftex,bookmarks,breaklinks,bookmarksopen,bookmarksnumbered,colorlinks,linkcolor=LINKCOL,linktocpage,citecolor=CITECOL,urlcolor=URLCOL,pdfpagemode=UseOutline,pdftex]{hyperref}

\usepackage{fancyhdr}
\fancypagestyle{titlepage}
{\chead{file: \jobname}\rhead{Total pages: \pageref{page:end}}}

\usepackage{booktabs}
\usepackage{natbib}

\definecolor{TITLECOL}{rgb}{0.1,0.2,0.7} 
\definecolor{SECOL}{rgb}{0.1,0.2,0.7} 
\definecolor{CONTENTSCOL}{rgb}{0.1,0.2,0.7} 
\definecolor{SSECOL}{rgb}{0.25,0,0.48} 
\definecolor{SSSECOL}{rgb}{0.2,0.08,0.53} 
\definecolor{FINCOL}{rgb}{0.01,0.3,0.07} 

\def\coloredtitle#1{\title{\textcolor{TITLECOL}{#1}}} 
\def\coloredauthor#1{\author{\textcolor{CITECOL}{#1}}} 

\definecolor{URLCOL}{rgb}{0,0.17,0.43} 
\definecolor{LINKCOL}{rgb}{0.05,0.4,0} 
\definecolor{CITECOL}{rgb}{0.35,0,0.48} 

\def\sss{\scriptscriptstyle\rm}

\def\bea{\begin{eqnarray}}
\def\eea{\end{eqnarray}}
\def\ben{\begin{equation}}
\def\een{\end{equation}}
\def\benu{\begin{enumerate}}
\def\enu{\end{enumerate}}

\def\bei{\begin{itemize}}
\def\eei{\end{itemize}}
\def\beit{\begin{itemize}}
\def\eit{\end{itemize}}
\def\benu{\begin{enumerate}}
\def\enu{\end{enumerate}}

\def\br{{\bf r}}

\def\half{\frac{1}{2}}

\def\x{_{\sss X}}

\def\s{_{\sss S}}

\def\xc{_{\sss XC}}

\def\LDA{^{\rm LDA}}

\def\TF{^{\rm TF}}

\def\unif{^{\rm unif}}
\def\up{_\uparrow}
\def\dn{_\downarrow}

\def\n{n}

\begin{document}

\coloredtitle{
Testing and using the Lewin-Lieb bounds in density functional theory
}
\coloredauthor{David V. Feinblum}
\affiliation{Department of Chemistry, University of California, Irvine, CA 92697}
\coloredauthor{John Kenison}
\affiliation{Department of Physics and Astronomy, University of California, Irvine, CA 92697}
\coloredauthor{Kieron Burke}
\affiliation{Department of Chemistry, University of California, Irvine, CA 92697}
\affiliation{Department of Physics and Astronomy, University of California, Irvine, CA 92697}
\date{\today}

\begin{abstract}
Lewin and Lieb have
recently proven several new bounds on the exchange-correlation energy that complement the
Lieb-Oxford bound.  We test these bounds for atoms, for slowly-varying gases, and
for Hooke's atom, finding them usually less strict than
the Lieb-Oxford bound.  However, we also show that,
if a generalized gradient approximation (GGA) is to guarantee satisfaction
of the new bounds for all densities, new restrictions on the 
the exchange-correlation enhancement factor are implied.
\end{abstract}

\maketitle


\section{Introduction}
The Lieb-Oxford (LO) bound \cite{LO81} is a cornerstone of
exact conditions in modern density functional theory.\cite{FNM03}  
Rigorously proven for non-relativistic quantum systems, the LO bound provides a strict upper bound on the magnitude of the
exchange-correlation energy, $E\xc$, of any system
relative to a simple integral over its density.   
The constant in the LO bound was built into
the construction of the PBE generalized gradient 
approximation\cite{PBE96}, one of the most popular
approximations in use in density functional theory (DFT) today.\cite{B14}  
The use of the bound to construct approximations
remains somewhat controversial, as most systems' $E\xc$ does not
come close to reaching this bound.\cite{OC07}

Recently, Lewin and Lieb \cite{LL14} have 
proven several alternative forms for the bound
that are distinct from the original LO bound. 
In each, some fraction of the density integral is traded for an integral
over a density gradient. Thus, the new bounds are tighter for uniform
and slowly varying gases, and could be hoped to be tighter for real systems.
If so, they would be more useful than the LO bound 
in construction and testing of approximate density functionals. 
We show below that they are not tighter for atoms or for Hooke's atom
(two electrons in a parabolic well).  However, they do lead to new restrictions
on the enhancement factor in GGAs that are constructed to guarantee 
satisfaction of the bounds for all possible densities.

To begin, the LO bound can be written as
\ben
E\xc \ge - C_{LO}  \int d^3r\, \n^{4/3}(\br)
\label{LO}
\een
where $C_{LO}$ is a constant that Lieb and Oxford\cite{LO81} showed is no larger than 1.68
(Chan and Handy showed it to be no larger than 1.6358).\cite{CH99}  
For simplicity we define the following density integrals:
\ben
I_0= \int d^3r\, \n^{4/3}(\br),
\label{I0}
\een
\ben
I_1 = \int d^3r\, | \nabla \n(\br)|,
\label{I1}
\een
\ben
I_2 = \int d^3r\, | \nabla \n^{1/3}(\br)|^2.
\label{I2}
\een
Two new families of bounds are derived by Lewin and Lieb\cite{LL14}:
\ben
U\xc \geq -C_{LL}\, I_0 -\alpha\, I_0 - c_p\, I_p /\alpha^{k-1}, ~~~p=1,2
\label{LLind}
\een
where $C_{LL}=3(9\pi/2)^{1/3}/5 \approx 1.4508$, 
$c_1=1.206\times 10^{-3}$, $c_2=0.2097$, $k=5-p$, and $\alpha$ is any positive number.
Here $U\xc$ is the potential energy contribution to the exchange-correlation energy.

We can convert these to a family of conditions on $E\xc$ with several simple steps.
We utilize the adiabatic connection formula\cite{LP75,GL76} in terms of the scaled density:
\ben
E\xc = \int_0^1 d\lambda \lambda\, U\xc[\n_{1/\lambda}],
\een
where $\n_\gamma (\br) = \gamma^3\, \n(\gamma\br)$ is the density 
scaled uniformly\cite{LP85,L91} 
by positive constant $\gamma$. Examining each of the integrals in the LL bounds, we
find
\ben
I_p [\n_\gamma] = \gamma\, I_p[\n],
\een
so that applying the bounds to every value of $\lambda$ between 0 and 1 yields
a bound on the DFT exchange-correlation energy directly:
\ben
E\xc \geq -C_{LL}\, I_0 -\alpha\, I_0 - c_p\, I_p /\alpha^{k-1}, ~~~p=1,2.
\label{LL}
\een

There is a Faustian dilemma when it comes to the value of $\alpha$:  A very small value
makes the first term smaller in magnitude than that of the LO bound, but
increases the of the gradient additions.  A very large value will
make those additions negligible, but also make the first term larger than that
of LO bound.   
Choosing $\alpha$ in each case to maximize the right-hand-side,
Lewin and Lieb find
\ben
E\xc \geq -C_{LL}\, I_0 - \tilde c_p\, I_p^{1/k}\, I_0^{1-1/k}  ~~~p=1,2
\label{LL}
\een
where 
\ben
\tilde c_p = \frac{k}{k-1}\, ((k-1) c_p)^{1/k}.
\label{LL123}
\een
This yields $\tilde c_1=4 (3 c_1)^{1/4}/3 \approx 0.3207$ and 
$\tilde c_2=3 (2 c_2)^{1/3}/2 \approx 1.1227$.
Lewin and Lieb report a third bound by combining the $p=2$ case with 
the Schwarz inequality:
\ben
E\xc \geq -C_{LL}\, I_0 - \tilde c_3\, I_2^{1/8}\, I_0^{7/8}
\label{LL3}
\een
where $\tilde c_3 = ((3^{1/4} c_{1})^{2/5})(c_{2}^{3/5}) \approx 0.7650$.
We refer to these as the optimized LL bounds with LL1 and LL2 given by (9) with $p=1,2$ respectively. LL3 is given by (11).

\section{Lewin-Lieb Bounds for Spherically Symmetric Atoms}
To test each of these bounds, we performed calculations
using the non-relativistic atomic OEP code of Engel.\cite{ED99} 
We used the PBE functional\cite{PBE96} to find self-consistent atomic
densities, and evaluated all $I_p$.  
We did this for a simple subset of atoms
for which highly accurate correlation energies are available.\cite{MT11}  
The results are fully converged with respect to the radial grid.
We use accurate $E\xc$ from Ref. \onlinecite{BCGP14}

In Table \ref{atoms}, we list the results.  We see immediately that, unfortunately, 
the new bounds are less restrictive than the current LO bound. 
Atoms have gradients that are sufficiently large as to
make the corrections larger than the density-integral term.
\begin{table}[htb]
\begin{tabular}{|c|c|cccc|ccc|}
\hline
$Z$ & $E_{XC}$ & LO & LL1 & LL2 & LL3 & $-C_{ll}I_{0}$ & $I_{1}$ & $I_{2}$ \\ \hline
1	&	-0.3125	&	-1.200	&	-1.395	&	-3.131	&	-2.159	&	-1.036	&	3.978	&	12.73	\\
2	&	-1.069	&	-1.991	&	-2.318	&	-4.514	&	-3.302	&	-1.719	&	6.739	&	10.99	\\
4	&	-2.758	&	-5.255	&	-6.095	&	-11.19	&	-8.401	&	-4.538	&	16.82	&	21.26	\\
10	&	-12.51	&	-25.01	&	-28.56	&	-43.09	&	-35.35	&	-21.60	&	62.18	&	31.64	\\
12	&	-16.47	&	-33.20	&	-37.83	&	-56.93	&	-46.80	&	-28.67	&	79.85	&	40.82	\\
18	&	-30.97	&	-63.36	&	-71.82	&	-101.2	&	-85.64	&	-54.72	&	139.5	&	49.76	\\
20	&	-36.11	&	-74.16	&	-83.97	&	-118.6	&	-100.3	&	-64.04	&	160.4	&	58.77	\\
30	&	-71.22	&	-149.1	&	-167.6	&	-220.0	&	-192.3	&	-128.8	&	284.1	&	68.21	\\
36	&	-95.79	&	-201.5	&	-225.9	&	-298.8	&	-256.0	&	-174.0	&	365.7	&	76.26	\\
38	&	-104.0	&	-219.2	&	-245.5	&	-316.2	&	-278.9	&	-189.3	&	393.2	&	84.90	\\
48	&	-151.7	&	-321.9	&	-359.3	&	-446.9	&	-400.2	&	-278.0	&	542.4	&	92.74	\\
54	&	-182.2	&	-388.0	&	-432.4	&	-531.7	&	-478.7	&	-335.1	&	635.9	&	100.6	\\
56	&	-192.4	&	-410.1	&	-456.8	&	-563.8	&	-506.9	&	-354.2	&	667.1	&	109.3	\\
70	&	-281.1	&	-603.6	&	-669.6	&	-799.6	&	-729.4	&	-521.3	&	911.6	&	117.9	\\
80	&	-350.5	&	-755.4	&	-836.1	&	-981.7	&	-902.0	&	-652.3	&	1096	&	124.9	\\
86	&	-393.0	&	-848.5	&	-938.1	&	-1096	&	-1009	&	-732.7	&	1210	&	132.5	\\
88	&	-405.2	&	-879.3	&	-972.0	&	-1139	&	-1048	&	-759.3	&	1246	&	141.0	\\

\hline
\end{tabular}
\caption{Exchange-correlation energies for neutral spherical atoms, bounds, and integrals.}
\label{atoms}
\end{table}

\begin{figure}[htb]
\includegraphics[width=0.4\textwidth]{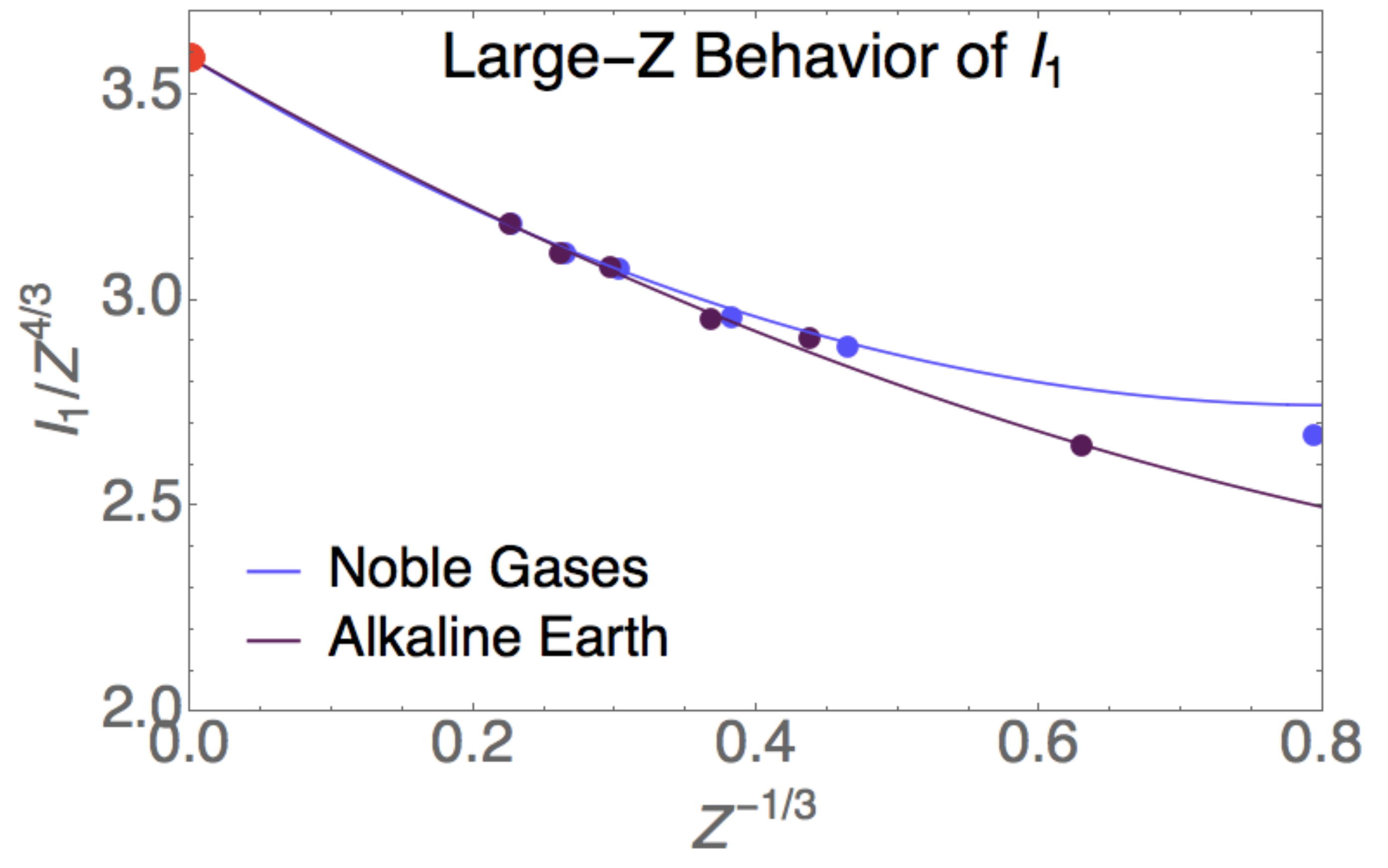}
\caption{$I_1$ (see text) for noble gas and alkaline earth atoms. The red dot indicates the limiting value as $Z \to \infty$}
\label{I1}
\end{figure}
\begin{figure}[htb]
\includegraphics[width=0.4\textwidth]{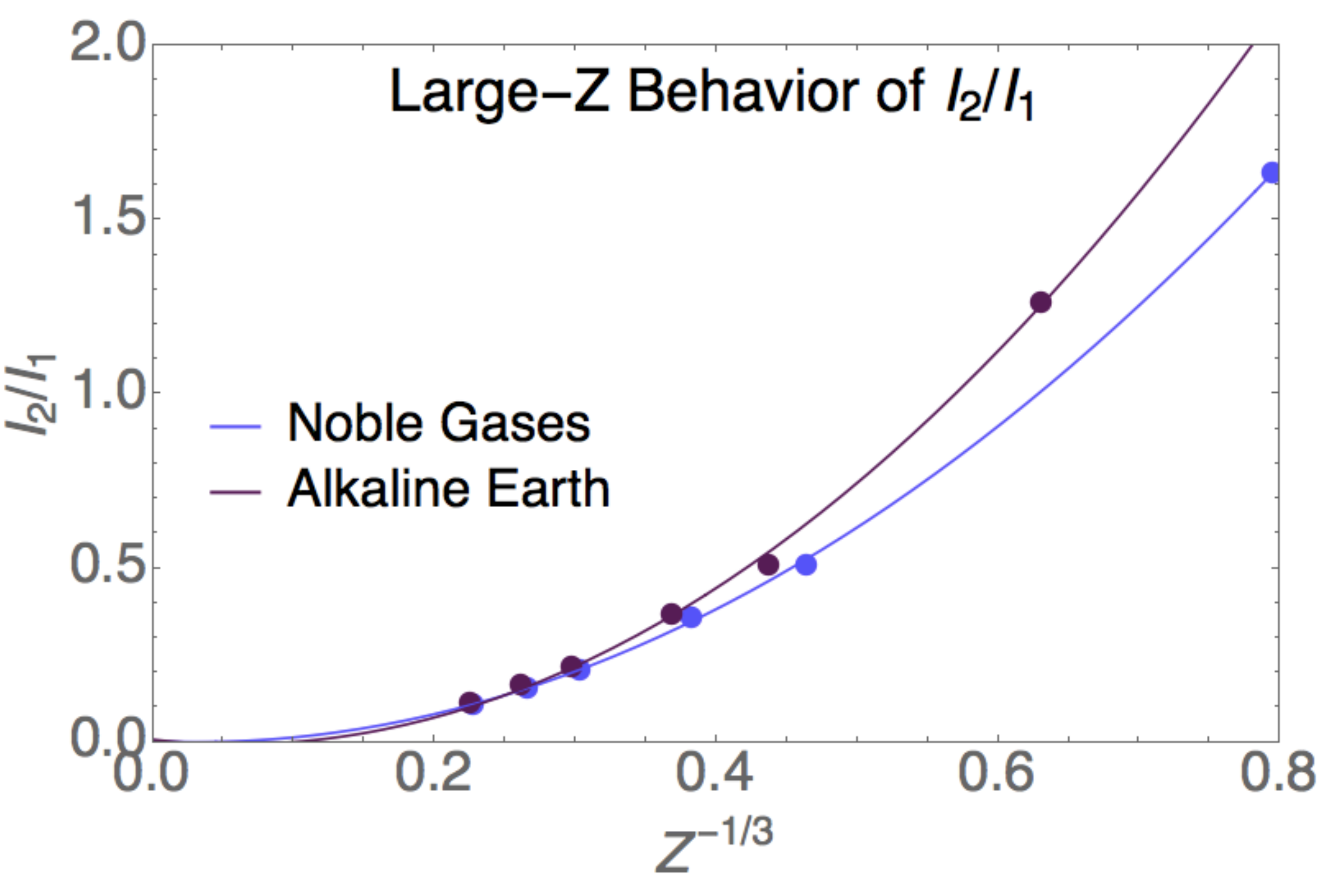}
\caption{$I_2/I_1$ ratio (see text) for noble gas and alkaline earth atoms.}
\label{I2}
\end{figure}
To be sure that {\em no} atom behaves differently, we examine the large-$Z$ limit, where
Thomas-Fermi theory applies.\cite{LS73,LCPB09}  It has recently been shown\cite{BCGP14}
that
\ben
E\xc \to -C\x Z^{5/3} - A\, Z\, \ln Z + B\xc\, Z+...
\een
for atoms, where $C\x=0.2201$, $A=0.020..$, and $B\xc \approx 0.039$.
The dominant term, which is an exchange contribution, was proven by Schwinger\cite{S81},
and can be easily calculated by inserting the TF density\cite{LCPB09} into the local approximation for 
$E\x$.  Since 
\ben
E\x\LDA = - A\x\, I_0
\een
where $A\x=\frac{3}{4}\left(\frac{3}{\pi}\right)^{1/3} \approx 0.738$, this easily satisfies all bounds, including any LL bound with the gradient
terms ignored.  In Fig. \ref{I1}, we plot $I_1$ as a function of $Z^{-1/3}$,
to show that it approaches its large $Z$ limit, which can be extracted from the TF density:
\ben
I_1[\n\TF] = d_1\TF\, Z^{4/3}
\een
where we find $d_1=3.58749$.   In Fig. \ref{I2}, we plot the ratio $I_2/I_1$, showing that, although $I_2$
diverges in TF theory (as noted by Lewin and Lieb), it appears to vanish relative to $I_1$ in this limit.
Thus all the additions in the LL bounds become relatively small in this limit, and no
change in behavior occurs. As $Z \to \infty$, the LL1 bound eventually becomes more restrictive than the LO bound, but only at unrealistically \footnote{$Z_{c} = \left[\frac{d_{1}}{C_{x}} \left(\frac{C_{1}}{C_{LO}-C_{LL}}\right)^{4}\right]^{3} \approx 91493$} large values of Z.

\section{Hooke's Atom and The Slowly-Varying Electron Gas}

We also performed calculations on the model system of two electrons in a harmonic potential, 
the Hooke's atom.\cite{T93} One might imagine that, for higher or lower densities, 
the bounds might tighten, or their order reverse, given the
different external potential.
We report three distinct results.  For $k\to\infty$, where $k$ is the spring
constant, the density becomes large, and $E\xc\to E\x$.  All energies and
integrals scale as $\omega^{1/2}$, where $\omega={\sqrt{k}}$.  The first
line of Table \ref{hooke} shows the results, which are analogous to those
of the two-electron ions (with different constants).  The order of the bounds
remains the same as in Table \ref{atoms}.  In the next line, we report actual
energies for the largest value of $k$ for which there exists an analytic solution, $k=1/4.$ Again we see the same behavior.

The most interesting case is the low-density limit, $k\to 0$.  In this limit, the
kinetic energy becomes negligible, and the electrons arrange themselves to minimize
the potential energy, on opposite sides of the center. This regime provides a system where correlation energy becomes comparable to exchange energy.
The third line of Table \ref{hooke} shows that none of the bounds is
tight in this limit (the XC energy vanishes relative to any of them)
and that the LL bounds diverge relative to the LO bound.
\def\kp{{k^{1/36}}}
\begin{table}[htb]
\begin{tabular}{|c|c|c|c|c|c|c|}
\hline
$k$ & $scale$ & $E_{xc}$ & $LO$ & $LL1$ & $LL2$ & $LL3$ \\ \hline
$\infty$& $k^{1/4}$ & -1.37   & -1.5513 &-1.7879  & -3.1804  & -2.4255 \\
$\frac{1}{4}$  &  1   &  -.554 & -1.0031  &-1.1558 & -2.0682 & -1.5740\\
$0$ & $k^{11/36}$ & -.0042 $k^{1/36}$
& -1.85
& -1.6-$\frac{.44}{k^{1/24}}$
& -1.6-$\frac{3.34}{k^{1/6}} $
&-1.6-$\frac{1.8}{k^{19/144}} $\\
\hline
\end{tabular}
\caption{Hooke's atom (two electrons in a harmonic potential) ranging over
all values of the spring constant, $k$.}
\label{hooke}
\end{table}

\section{Effects on $F\xc$}
 
In the rest of this paper, we show how the LL bounds can be used to 
derive interesting and new restrictions on the enhancement factor of generalized 
gradient approximations (GGA's).
Begin with the definition of the enhancement factor for a GGA for spin unpolarized
systems:
\ben
E\xc^{GGA} = \int d^3r\, e\x\unif(\n(\br))\, F\xc(r\s(\br),s(\br)),
\een
where $e\x\unif(\n)= -A\x \n^{4/3}$ is the exchange energy density of a
spin-unpolarized uniform gas, $r\s= (3/(4\pi\n))^{1/3}$ is the local
Wigner-Seitz radius, and $s= |\nabla \n|/(2 k_F \n)$ is the (exchange) dimensionless
measure of the gradient, where $k_F = (3 \pi^2 \n)^{1/3}$ is the local
Fermi wavevector.
Most famously, the PBE approximation was constructed to ensure it satisfies the
LO bound for any density. A sufficient condition to guarantee this is
\ben
F\x (s) \leq 1.804.
\label{FxLO}
\een

Here we digress slightly, to correct a popular misconception in the literature.\cite{VRLM12}
The LO bound applies to the XC energy. There is no unique choice of XC energy density, 
and more than one
choice was used in the derivation of PBE.\cite{BPW97}  Thus the enhancement 
factor in PBE should not (and does not) correspond to {\em any} choice of 
energy density.  No bound has ever been defined, much less proven, for a specific energy
density.
As others\cite{OC07}  and Table \ref{atoms} have shown, real systems do not come close to saturating 
the LO bound.  In fact, the B88 exchange functional does not satisfy Eq. (\ref{FxLO})
due to the logarithmic dependence on $s$.  But
for the present purposes, B88 gives exchange energies almost identical to 
PBE and very close to exact exchange energies for atoms. 
While one can design densities that cause B88 to violate the LO bound, 
they look nothing like densities of real systems.\cite{PRSB14}

Now we apply the logic of PBE to the LL bounds.  We wish to find 
conditions on the enhancement factor that guarantee
satisfaction of those bounds for all possible densities. In this context, the
optimum bounds are not useful, since they contain denominators different from the local term. 
Dividing each term of Eq. (8) by $e\x\unif$, we find that
\ben
F\xc \leq {\tilde C}_{LL} + A\x^{-1}(\alpha + c'_p\, s^p/\alpha^{k-1})
\een
is a sufficient condition to ensure satisfaction of the LL bounds for any density, with $\tilde C_{LL}=C_{LL}/A\x= 6 (2\pi/3)^{2/3}/5=1.9643...$, and 
\ben
c'_p = 2\, (4-p)\, \left( \frac{\pi}{3} \right)^{2p/3} c_{p} .
\een
We now find the most restrictive value of $\alpha$ for each value of $s$, to yield
\ben
F\xc \leq {\tilde C}_{LL} + \tilde d_p\, s^{p/(5-p)},
\een
where
\ben
\tilde d_p = \frac{k}{k-1} \left( (k-1)\, c'_p \right)^{1/k}.
\een
Writing these out explicitly yields
\ben
F\xc \leq 1.9643 + 2.76755\, s^{1/4},~~~~~~~~~(\text{LL1})
\een
and
\ben
F\xc \leq 1.9643 + 3.06212\, s^{2/3},~~~~~~~~~(\text{LL2})
\een
in contrast to the LO bound
\ben
F\xc \leq 2.273.~~~~~~~~~~(\text{LO})
\een
\begin{center}
\begin{figure}[htb]
\includegraphics[width=0.47\textwidth]{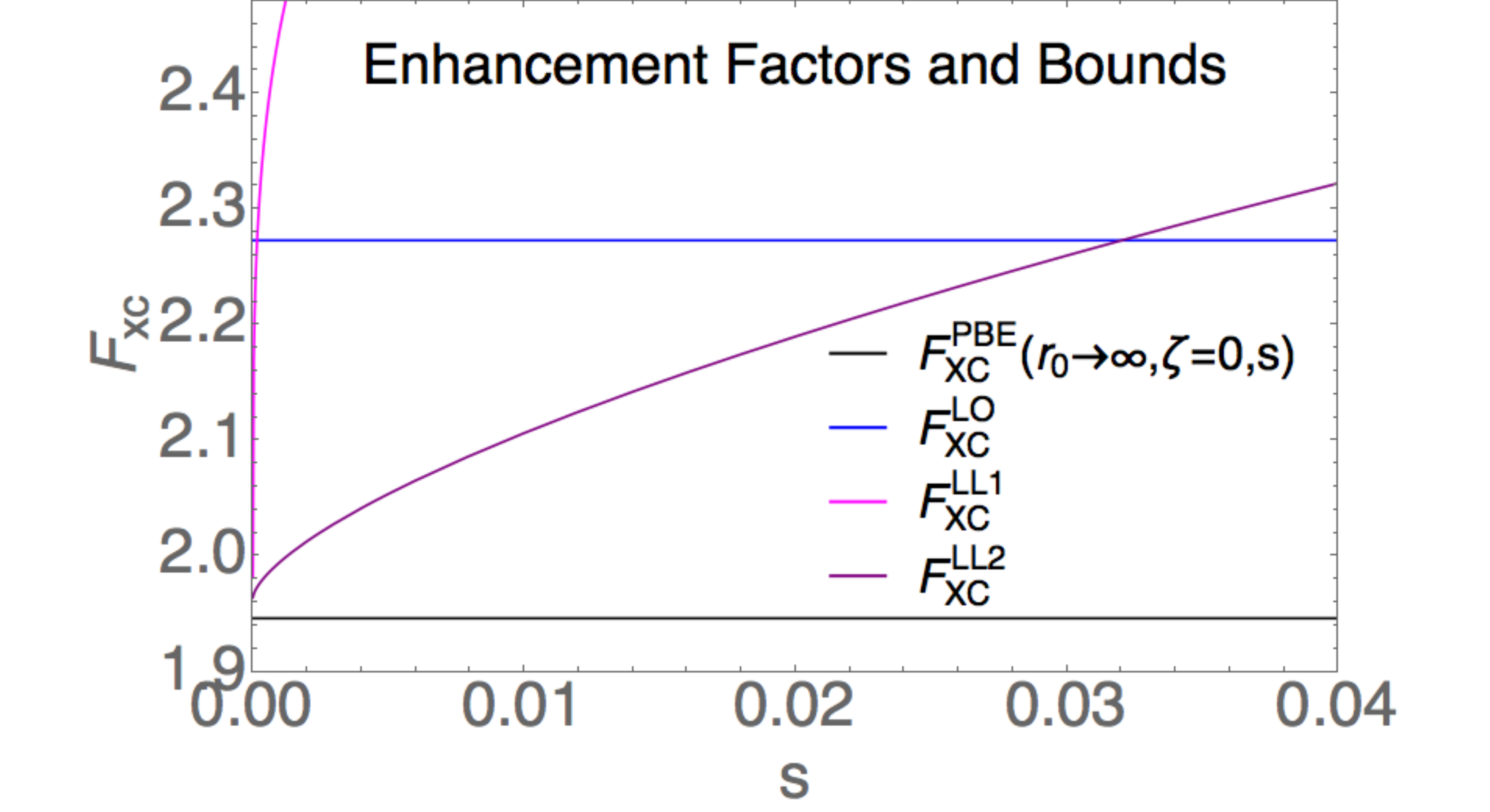}
\caption{$F\xc$ for LL and LO compared with PBE density for small $s$.}
\label{Fxc}
\end{figure}
\end{center}
In Fig 3, we plot all three bounds and see that for small values of $s$, the LL2 bound is tighter than the LO bound. We therefore define the new Lieb-Oxford-Lewin (LOL) bound to be the LL2 bound for small $s$ and the old LO bound otherwise.
This new bound is more restrictive than the LO bound when $s$ is less than 0.032. The existence of a tighter bound on the enhancement factor for all s has been empirically suggested.\cite{RPCP09} We also plot the PBE enhancement factor, showing that it satisfies the LOL bound, but is much closer for small $s$ than the old LO bound. 

Spin-polarization is handled different for exchange than for correlation.
For exchange-correlation together, it does not raise $F\xc$ beyond its maximum
for unpolarized systems, as that is achieved in the low-density limit, which
is independent of spin.  But in the opposite, high-density limit, exchange
dominates, and its spin-dependence is determined by the exact spin-scaling
relation for exchange:
\ben
E\x [\n\up,\n\dn] = \half( E\x[2\n\up,0]+E\x[0,2\n\dn] )
\een
which implies
\ben
F\x^{pol} (s) = 2^{1/3}\, F\x^{unpol} (s)
\een
Thus $F\x^{unpol} < \tilde C_{LO}/2^{1/3}$ to ensure $F\x^{pol}$ satisfies the LO
bound, which is the origin of Eq. (\ref{FxLO}).

\section{Conclusion}
In summary, we have tested the optimum LL bounds (which have already been applied to DFT by other groups\cite{CT14}) for a variety of simple systems,
finding they are less restrictive for those systems than the LO bound. 
However, the LL bounds are clearly more restrictive for a uniform gas, and the family
of bounds that the LL bounds come from can be used to place limits on the
enhancement factor of GGA's. With this in mind, we constructed the combined
LOL bound and we recommend that the LOL bound be used whenever relevant for all future 
functional development and testing of GGAs.

\section{Acknowledgements}
We thank Mathieu Lewin and Elliott Lieb for bringing their new bounds to our
attention, and
Eberhard Engel for developing the OPMKS atom code.
This work was supported by NSF under grant CHE-1112442.

\label{page:end}
\bibliography{Master,dave}

\begin{thebibliography}{24}%
\makeatletter
\providecommand \@ifxundefined [1]{%
 \@ifx{#1\undefined}
}%
\providecommand \@ifnum [1]{%
 \ifnum #1\expandafter \@firstoftwo
 \else \expandafter \@secondoftwo
 \fi
}%
\providecommand \@ifx [1]{%
 \ifx #1\expandafter \@firstoftwo
 \else \expandafter \@secondoftwo
 \fi
}%
\providecommand \natexlab [1]{#1}%
\providecommand \enquote  [1]{``#1''}%
\providecommand \bibnamefont  [1]{#1}%
\providecommand \bibfnamefont [1]{#1}%
\providecommand \citenamefont [1]{#1}%
\providecommand \href@noop [0]{\@secondoftwo}%
\providecommand \href [0]{\begingroup \@sanitize@url \@href}%
\providecommand \@href[1]{\@@startlink{#1}\@@href}%
\providecommand \@@href[1]{\endgroup#1\@@endlink}%
\providecommand \@sanitize@url [0]{\catcode `\\12\catcode `\$12\catcode
  `\&12\catcode `\#12\catcode `\^12\catcode `\_12\catcode `\%12\relax}%
\providecommand \@@startlink[1]{}%
\providecommand \@@endlink[0]{}%
\providecommand \url  [0]{\begingroup\@sanitize@url \@url }%
\providecommand \@url [1]{\endgroup\@href {#1}{\urlprefix }}%
\providecommand \urlprefix  [0]{URL }%
\providecommand \Eprint [0]{\href }%
\providecommand \doibase [0]{http://dx.doi.org/}%
\providecommand \selectlanguage [0]{\@gobble}%
\providecommand \bibinfo  [0]{\@secondoftwo}%
\providecommand \bibfield  [0]{\@secondoftwo}%
\providecommand \translation [1]{[#1]}%
\providecommand \BibitemOpen [0]{}%
\providecommand \bibitemStop [0]{}%
\providecommand \bibitemNoStop [0]{.\EOS\space}%
\providecommand \EOS [0]{\spacefactor3000\relax}%
\providecommand \BibitemShut  [1]{\csname bibitem#1\endcsname}%
\let\auto@bib@innerbib\@empty
\bibitem [{\citenamefont {Lieb}\ and\ \citenamefont {Oxford}(1981)}]{LO81}%
  \BibitemOpen
  \bibfield  {author} {\bibinfo {author} {\bibfnamefont {Elliott~H.}\
  \bibnamefont {Lieb}}\ and\ \bibinfo {author} {\bibfnamefont {Stephen}\
  \bibnamefont {Oxford}},\ }\bibfield  {title} {\enquote {\bibinfo {title}
  {Improved lower bound on the indirect coulomb energy},}\ }\href {\doibase
  10.1002/qua.560190306} {\bibfield  {journal} {\bibinfo  {journal}
  {International Journal of Quantum Chemistry}\ }\textbf {\bibinfo {volume}
  {19}},\ \bibinfo {pages} {427--439} (\bibinfo {year} {1981})}\BibitemShut
  {NoStop}%
\bibitem [{\citenamefont {Fiolhais}\ \emph {et~al.}(2003)\citenamefont
  {Fiolhais}, \citenamefont {Nogueira},\ and\ \citenamefont {Marques}}]{FNM03}%
  \BibitemOpen
  \bibfield  {author} {\bibinfo {author} {\bibfnamefont {Carlos}\ \bibnamefont
  {Fiolhais}}, \bibinfo {author} {\bibfnamefont {F.}~\bibnamefont {Nogueira}},
  \ and\ \bibinfo {author} {\bibfnamefont {M.}~\bibnamefont {Marques}},\ }\href
  {http://books.google.com/books?id=mX793GABep8C&dq=A+Primer+in+Density+Functional+Theory&printsec=frontcover&source=bn&hl=en&ei=IVt8TNDSEYeCsQOa0bCEBw&sa=X&oi=book_result&ct=result&resnum=4&ved=0CCYQ6AEwAw\#v=onepage&q&f=false}
  {\emph {\bibinfo {title} {A Primer in Density Functional Theory}}}\ (\bibinfo
   {publisher} {Springer-Verlag},\ \bibinfo {address} {New York},\ \bibinfo
  {year} {2003})\BibitemShut {NoStop}%
\bibitem [{\citenamefont {Perdew}\ \emph {et~al.}(1996)\citenamefont {Perdew},
  \citenamefont {Burke},\ and\ \citenamefont {Ernzerhof}}]{PBE96}%
  \BibitemOpen
  \bibfield  {author} {\bibinfo {author} {\bibfnamefont {John~P.}\ \bibnamefont
  {Perdew}}, \bibinfo {author} {\bibfnamefont {Kieron}\ \bibnamefont {Burke}},
  \ and\ \bibinfo {author} {\bibfnamefont {Matthias}\ \bibnamefont
  {Ernzerhof}},\ }\bibfield  {title} {\enquote {\bibinfo {title} {Generalized
  gradient approximation made simple},}\ }\href {\doibase
  10.1103/PhysRevLett.77.3865} {\bibfield  {journal} {\bibinfo  {journal}
  {Phys. Rev. Lett.}\ }\textbf {\bibinfo {volume} {77}},\ \bibinfo {pages}
  {3865--3868} (\bibinfo {year} {1996})},\ \bibinfo {note} {{\it ibid.} {\bf
  78}, 1396(E) (1997)}\BibitemShut {NoStop}%
\bibitem [{\citenamefont {Becke}({2014})}]{B14}%
  \BibitemOpen
  \bibfield  {author} {\bibinfo {author} {\bibfnamefont {Axel~D.}\ \bibnamefont
  {Becke}},\ }\bibfield  {title} {\enquote {\bibinfo {title} {{Perspective:
  Fifty years of density-functional theory in chemical physics}},}\ }\href
  {\doibase {10.1063/1.4869598}} {\bibfield  {journal} {\bibinfo  {journal}
  {{JOURNAL OF CHEMICAL PHYSICS}}\ }\textbf {\bibinfo {volume} {{140}}}
  (\bibinfo {year} {{2014}}),\ {10.1063/1.4869598}}\BibitemShut {NoStop}%
\bibitem [{\citenamefont {Odashima}\ and\ \citenamefont
  {Capelle}(2007)}]{OC07}%
  \BibitemOpen
  \bibfield  {author} {\bibinfo {author} {\bibfnamefont {Mariana~M.}\
  \bibnamefont {Odashima}}\ and\ \bibinfo {author} {\bibfnamefont
  {K.}~\bibnamefont {Capelle}},\ }\bibfield  {title} {\enquote {\bibinfo
  {title} {How tight is the lieb-oxford bound?}}\ }\href {\doibase
  http://dx.doi.org/10.1063/1.2759202} {\bibfield  {journal} {\bibinfo
  {journal} {Journal of Chemical Physics}\ } (\bibinfo {year} {2007}),\
  http://dx.doi.org/10.1063/1.2759202}\BibitemShut {NoStop}%
\bibitem [{\citenamefont {Lewin}\ and\ \citenamefont {H~Lieb}(2014)}]{LL14}%
  \BibitemOpen
  \bibfield  {author} {\bibinfo {author} {\bibfnamefont {Mathieu}\ \bibnamefont
  {Lewin}}\ and\ \bibinfo {author} {\bibfnamefont {Elliott}\ \bibnamefont
  {H~Lieb}},\ }\bibfield  {title} {\enquote {\bibinfo {title} {Improved
  lieb-oxford exchange-correlation inequality with gradient correction},}\
  }\href@noop {} {\bibfield  {journal} {\bibinfo  {journal} {arXiv preprint
  arXiv:1408.3358v3}\ } (\bibinfo {year} {2014})}\BibitemShut {NoStop}%
\bibitem [{\citenamefont {Chan}\ and\ \citenamefont {Handy}(1999)}]{CH99}%
  \BibitemOpen
  \bibfield  {author} {\bibinfo {author} {\bibfnamefont {G.K-L.}\ \bibnamefont
  {Chan}}\ and\ \bibinfo {author} {\bibfnamefont {N.C.}\ \bibnamefont
  {Handy}},\ }\bibfield  {title} {\enquote {\bibinfo {title} {Optimized
  lieb-oxford bound for the exchange-correlation energy},}\ }\href@noop {}
  {\bibfield  {journal} {\bibinfo  {journal} {Phys. Rev. A}\ }\textbf {\bibinfo
  {volume} {59}},\ \bibinfo {pages} {3075} (\bibinfo {year}
  {1999})}\BibitemShut {NoStop}%
\bibitem [{\citenamefont {Langreth}\ and\ \citenamefont {Perdew}(1975)}]{LP75}%
  \BibitemOpen
  \bibfield  {author} {\bibinfo {author} {\bibfnamefont {D.C.}\ \bibnamefont
  {Langreth}}\ and\ \bibinfo {author} {\bibfnamefont {J.P.}\ \bibnamefont
  {Perdew}},\ }\bibfield  {title} {\enquote {\bibinfo {title} {The
  exchange-correlation energy of a metallic surface},}\ }\href@noop {}
  {\bibfield  {journal} {\bibinfo  {journal} {Solid State Commun.}\ }\textbf
  {\bibinfo {volume} {17}},\ \bibinfo {pages} {1425} (\bibinfo {year}
  {1975})}\BibitemShut {NoStop}%
\bibitem [{\citenamefont {Gunnarsson}\ and\ \citenamefont
  {Lundqvist}(1976)}]{GL76}%
  \BibitemOpen
  \bibfield  {author} {\bibinfo {author} {\bibfnamefont {O.}~\bibnamefont
  {Gunnarsson}}\ and\ \bibinfo {author} {\bibfnamefont {B.I.}\ \bibnamefont
  {Lundqvist}},\ }\bibfield  {title} {\enquote {\bibinfo {title} {Exchange and
  correlation in atoms, molecules, and solids by the spin-density-functional
  formalism},}\ }\href@noop {} {\bibfield  {journal} {\bibinfo  {journal}
  {Phys. Rev. B}\ }\textbf {\bibinfo {volume} {13}},\ \bibinfo {pages} {4274}
  (\bibinfo {year} {1976})}\BibitemShut {NoStop}%
\bibitem [{\citenamefont {Levy}\ and\ \citenamefont {Perdew}(1985)}]{LP85}%
  \BibitemOpen
  \bibfield  {author} {\bibinfo {author} {\bibfnamefont {M.}~\bibnamefont
  {Levy}}\ and\ \bibinfo {author} {\bibfnamefont {J.P.}\ \bibnamefont
  {Perdew}},\ }\bibfield  {title} {\enquote {\bibinfo {title}
  {Hellmann-feynman, virial, and scaling requisites for the exact universal
  density functionals. shape of the correlation potential and diamagnetic
  susceptibility for atoms},}\ }\href {\doibase 10.1103/PhysRevA.32.2010}
  {\bibfield  {journal} {\bibinfo  {journal} {Phys. Rev. A}\ }\textbf {\bibinfo
  {volume} {32}},\ \bibinfo {pages} {2010} (\bibinfo {year}
  {1985})}\BibitemShut {NoStop}%
\bibitem [{\citenamefont {Levy}(1991)}]{L91}%
  \BibitemOpen
  \bibfield  {author} {\bibinfo {author} {\bibfnamefont {M.}~\bibnamefont
  {Levy}},\ }\bibfield  {title} {\enquote {\bibinfo {title} {Density-functional
  exchange-correlation through coordinate scaling in adiabatic connection and
  correlation hole},}\ }\href@noop {} {\bibfield  {journal} {\bibinfo
  {journal} {Phys. Rev. A}\ }\textbf {\bibinfo {volume} {43}},\ \bibinfo
  {pages} {4637} (\bibinfo {year} {1991})}\BibitemShut {NoStop}%
\bibitem [{\citenamefont {Engel}\ and\ \citenamefont {Dreizler}(1999)}]{ED99}%
  \BibitemOpen
  \bibfield  {author} {\bibinfo {author} {\bibfnamefont {E.}~\bibnamefont
  {Engel}}\ and\ \bibinfo {author} {\bibfnamefont {R.~M.}\ \bibnamefont
  {Dreizler}},\ }\bibfield  {title} {\enquote {\bibinfo {title} {From explicit
  to implicit density functionals},}\ }\href {\doibase
  10.1002/(SICI)1096-987X(19990115)20:1<31::AID-JCC6>3.0.CO;2-P} {\bibfield
  {journal} {\bibinfo  {journal} {Journal of Computational Chemistry}\ }\textbf
  {\bibinfo {volume} {20}},\ \bibinfo {pages} {31--50} (\bibinfo {year}
  {1999})}\BibitemShut {NoStop}%
\bibitem [{\citenamefont {McCarthy}\ and\ \citenamefont
  {Thakkar}(2011)}]{MT11}%
  \BibitemOpen
  \bibfield  {author} {\bibinfo {author} {\bibfnamefont {Shane~P.}\
  \bibnamefont {McCarthy}}\ and\ \bibinfo {author} {\bibfnamefont {Ajit~J.}\
  \bibnamefont {Thakkar}},\ }\bibfield  {title} {\enquote {\bibinfo {title}
  {Accurate all-electron correlation energies for the closed-shell atoms from
  ar to rn and their relationship to the corresponding mp2 correlation
  energies},}\ }\href {\doibase http://dx.doi.org/10.1063/1.3547262} {\bibfield
   {journal} {\bibinfo  {journal} {The Journal of Chemical Physics}\ }\textbf
  {\bibinfo {volume} {134}},\ \bibinfo {eid} {044102} (\bibinfo {year}
  {2011})}\BibitemShut {NoStop}%
\bibitem [{\citenamefont {Burke}\ \emph {et~al.}(2014)\citenamefont {Burke},
  \citenamefont {Cancio}, \citenamefont {Gould},\ and\ \citenamefont
  {Pittalis}}]{BCGP14}%
  \BibitemOpen
  \bibfield  {author} {\bibinfo {author} {\bibfnamefont {Kieron}\ \bibnamefont
  {Burke}}, \bibinfo {author} {\bibfnamefont {Antonio}\ \bibnamefont {Cancio}},
  \bibinfo {author} {\bibfnamefont {Tim}\ \bibnamefont {Gould}}, \ and\
  \bibinfo {author} {\bibfnamefont {Stefano}\ \bibnamefont {Pittalis}},\
  }\bibfield  {title} {\enquote {\bibinfo {title} {Atomic correlation energies
  and the generalized gradient approximation},}\ }\href
  {http://arxiv.org/abs/1409.4834v1} {\bibfield  {journal} {\bibinfo  {journal}
  {submitted and arXiv:1409.4834v1}\ } (\bibinfo {year} {2014})}\BibitemShut
  {NoStop}%
\bibitem [{\citenamefont {Lieb}\ and\ \citenamefont {Simon}(1973)}]{LS73}%
  \BibitemOpen
  \bibfield  {author} {\bibinfo {author} {\bibfnamefont {E.H.}\ \bibnamefont
  {Lieb}}\ and\ \bibinfo {author} {\bibfnamefont {B.}~\bibnamefont {Simon}},\
  }\bibfield  {title} {\enquote {\bibinfo {title} {Thomas-fermi theory
  revisited},}\ }\href@noop {} {\bibfield  {journal} {\bibinfo  {journal}
  {Phys. Rev. Lett.}\ }\textbf {\bibinfo {volume} {31}},\ \bibinfo {pages}
  {681} (\bibinfo {year} {1973})}\BibitemShut {NoStop}%
\bibitem [{\citenamefont {Lee}\ \emph {et~al.}(2009)\citenamefont {Lee},
  \citenamefont {Constantin}, \citenamefont {Perdew},\ and\ \citenamefont
  {Burke}}]{LCPB09}%
  \BibitemOpen
  \bibfield  {author} {\bibinfo {author} {\bibfnamefont {Donghyung}\
  \bibnamefont {Lee}}, \bibinfo {author} {\bibfnamefont {Lucian~A.}\
  \bibnamefont {Constantin}}, \bibinfo {author} {\bibfnamefont {John~P.}\
  \bibnamefont {Perdew}}, \ and\ \bibinfo {author} {\bibfnamefont {Kieron}\
  \bibnamefont {Burke}},\ }\bibfield  {title} {\enquote {\bibinfo {title}
  {Condition on the kohn--sham kinetic energy and modern parametrization of the
  thomas--fermi density},}\ }\href {\doibase 10.1063/1.3059783} {\bibfield
  {journal} {\bibinfo  {journal} {J. Chem. Phys.}\ }\textbf {\bibinfo {volume}
  {130}},\ \bibinfo {eid} {034107} (\bibinfo {year} {2009})}\BibitemShut
  {NoStop}%
\bibitem [{\citenamefont {Schwinger}(1981)}]{S81}%
  \BibitemOpen
  \bibfield  {author} {\bibinfo {author} {\bibfnamefont {Julian}\ \bibnamefont
  {Schwinger}},\ }\bibfield  {title} {\enquote {\bibinfo {title} {Thomas-fermi
  model: The second correction},}\ }\href {\doibase 10.1103/PhysRevA.24.2353}
  {\bibfield  {journal} {\bibinfo  {journal} {Phys. Rev. A}\ }\textbf {\bibinfo
  {volume} {24}},\ \bibinfo {pages} {2353--2361} (\bibinfo {year}
  {1981})}\BibitemShut {NoStop}%
\bibitem [{Note1()}]{Note1}%
  \BibitemOpen
  \bibinfo {note} {$Z_{c} = \left [\protect \frac {d_{1}}{C_{x}} \left
  (\protect \frac {C_{1}}{C_{LO}-C_{LL}}\right )^{4}\right ]^{3} \approx
  91493$}\BibitemShut {NoStop}%
\bibitem [{\citenamefont {Taut}(1993)}]{T93}%
  \BibitemOpen
  \bibfield  {author} {\bibinfo {author} {\bibfnamefont {M.}~\bibnamefont
  {Taut}},\ }\bibfield  {title} {\enquote {\bibinfo {title} {Two electrons in
  an external oscillator potential: Particular analytic solutions of a coulomb
  correlation problem},}\ }\href@noop {} {\bibfield  {journal} {\bibinfo
  {journal} {Phys. Rev. A}\ }\textbf {\bibinfo {volume} {48}},\ \bibinfo
  {pages} {3561} (\bibinfo {year} {1993})}\BibitemShut {NoStop}%
\bibitem [{\citenamefont {Vilhena}\ \emph {et~al.}(2012)\citenamefont
  {Vilhena}, \citenamefont {R\"as\"anen}, \citenamefont {Lehtovaara},\ and\
  \citenamefont {Marques}}]{VRLM12}%
  \BibitemOpen
  \bibfield  {author} {\bibinfo {author} {\bibfnamefont {J.~G.}\ \bibnamefont
  {Vilhena}}, \bibinfo {author} {\bibfnamefont {E.}~\bibnamefont
  {R\"as\"anen}}, \bibinfo {author} {\bibfnamefont {L.}~\bibnamefont
  {Lehtovaara}}, \ and\ \bibinfo {author} {\bibfnamefont {M.~A.~L.}\
  \bibnamefont {Marques}},\ }\bibfield  {title} {\enquote {\bibinfo {title}
  {Violation of a local form of the lieb-oxford bound},}\ }\href {\doibase
  10.1103/PhysRevA.85.052514} {\bibfield  {journal} {\bibinfo  {journal} {Phys.
  Rev. A}\ }\textbf {\bibinfo {volume} {85}},\ \bibinfo {pages} {052514}
  (\bibinfo {year} {2012})}\BibitemShut {NoStop}%
\bibitem [{\citenamefont {Burke}\ \emph {et~al.}(1997)\citenamefont {Burke},
  \citenamefont {Perdew},\ and\ \citenamefont {Wang}}]{BPW97}%
  \BibitemOpen
  \bibfield  {author} {\bibinfo {author} {\bibfnamefont {Kieron}\ \bibnamefont
  {Burke}}, \bibinfo {author} {\bibfnamefont {John~P.}\ \bibnamefont {Perdew}},
  \ and\ \bibinfo {author} {\bibfnamefont {Y.}~\bibnamefont {Wang}},\ }\enquote
  {\bibinfo {title} {Derivation of a generalized gradient approximation: The
  pw91 density functional},}\ in\ \href@noop {} {\emph {\bibinfo {booktitle}
  {Electronic Density Functional Theory: Recent Progress and New
  Directions}}},\ \bibinfo {editor} {edited by\ \bibinfo {editor}
  {\bibfnamefont {J.~F.}\ \bibnamefont {Dobson}}, \bibinfo {editor}
  {\bibfnamefont {G.}~\bibnamefont {Vignale}}, \ and\ \bibinfo {editor}
  {\bibfnamefont {M.~P.}\ \bibnamefont {Das}}}\ (\bibinfo  {publisher}
  {Plenum},\ \bibinfo {address} {NY},\ \bibinfo {year} {1997})\ p.~\bibinfo
  {pages} {81}\BibitemShut {NoStop}%
\bibitem [{\citenamefont {Perdew}\ \emph {et~al.}(2014)\citenamefont {Perdew},
  \citenamefont {Ruzsinszky}, \citenamefont {Sun},\ and\ \citenamefont
  {Burke}}]{PRSB14}%
  \BibitemOpen
  \bibfield  {author} {\bibinfo {author} {\bibfnamefont {John~P.}\ \bibnamefont
  {Perdew}}, \bibinfo {author} {\bibfnamefont {Adrienn}\ \bibnamefont
  {Ruzsinszky}}, \bibinfo {author} {\bibfnamefont {Jianwei}\ \bibnamefont
  {Sun}}, \ and\ \bibinfo {author} {\bibfnamefont {Kieron}\ \bibnamefont
  {Burke}},\ }\bibfield  {title} {\enquote {\bibinfo {title} {Gedanken
  densities and exact constraints in density functional theory},}\ }\href@noop
  {} {\bibfield  {journal} {\bibinfo  {journal} {The Journal of Chemical
  Physics}\ }\textbf {\bibinfo {volume} {140}},\ \bibinfo {eid} {18} (\bibinfo
  {year} {2014})}\BibitemShut {NoStop}%
\bibitem [{\citenamefont {E.~Rasanen}\ and\ \citenamefont
  {Proetto}(2009)}]{RPCP09}%
  \BibitemOpen
  \bibfield  {author} {\bibinfo {author} {\bibfnamefont {K.~Capelle}\
  \bibnamefont {E.~Rasanen}, \bibfnamefont {S.~Pittalis}}\ and\ \bibinfo
  {author} {\bibfnamefont {C.~R.}\ \bibnamefont {Proetto}},\ }\bibfield
  {title} {\enquote {\bibinfo {title} {Lower bounds on the exchange-correlation
  energy in reduced dimensions},}\ }\href {\doibase
  http://dx.doi.org/10.1103/PhysRevLett.102.206406} {\bibfield  {journal}
  {\bibinfo  {journal} {Phys. Rev. Lett.}\ } (\bibinfo {year} {2009}),\
  http://dx.doi.org/10.1103/PhysRevLett.102.206406}\BibitemShut {NoStop}%
\bibitem [{\citenamefont {Constantin}\ and\ \citenamefont
  {Terentjevs}(2014)}]{CT14}%
  \BibitemOpen
  \bibfield  {author} {\bibinfo {author} {\bibfnamefont {Lucian~A.}\
  \bibnamefont {Constantin}}\ and\ \bibinfo {author} {\bibfnamefont
  {Aleksandrs}\ \bibnamefont {Terentjevs}},\ }\bibfield  {title} {\enquote
  {\bibinfo {title} {Gradient-dependent upper bound for the
  exchange-correlation energy and application to density functional theory},}\
  }\href {http://arxiv.org/pdf/1411.1579.pdf} {\bibfield  {journal} {\bibinfo
  {journal} {arXiv:1411.1579}\ } (\bibinfo {year} {2014})}\BibitemShut
  {NoStop}%
\end{thebibliography}%

\end{document}